\documentclass[epj]{svjour}

\usepackage{graphicx,array}
\usepackage{amssymb}

\DeclareGraphicsExtensions{.eps}

\def\sinc{\mathrm{sinc}}

\begin{document}

\title{Non-linear and quantum optics of a type II OPO containing a
  birefringent element Part 1: Classical operation}

\titlerunning{Non-linear and quantum optics of a type II OPO~$\dots$~(part~1)~:
  classical operation}

\author{L. Longchambon , J. Laurat , T. Coudreau
 \mail{coudreau@spectro.jussieu.fr} \and C. Fabre  }

\authorrunning{L. Longchambon \emph{et al.}}

\institute{Laboratoire Kastler Brossel, Case 74, UPMC, 4, Place
  Jussieu, 75252 Paris cedex 05, France }
\date{\today}

\abstract{We describe theoretically the main characteristics of the
  steady state regime of a type II Optical Parametric Oscillator (OPO)
  containing a birefringent plate. In such a device the signal and
  idler waves are at the same time linearly coupled by the plate and
  nonlinearly coupled by the $\chi^{(2)}$ crystal. This mixed coupling
  allows, in some well-defined range of the control parameters, a
  frequency degenerate operation as well as phase locking between the
  signal and idler modes. We describe here a complete model taking
  into account all possible effects in the system, \emph{i.e.} arbitrary rotation
   of the waveplate, non perfect phase matching, ring and linear
   cavities. This model is able to explain the detailed features of
   the experiments performed with this system.%
}

\maketitle

\section{Introduction}

\label{intro}

In a type II OPO, signal and idler fields of crossed polarizations
are generated when the pump exceeds a certain threshold. Energy
conservation requires that $\omega_0=\omega_1+\omega_2$, where
$\omega_0$, $\omega_1$ and $\omega_2$ are respectively the pump,
signal and idler frequencies. The precise values of the signal and
idler frequencies are set by the conditions of equal cavity
detunings and minimum oscillation threshold, which depend on the
value of the phase matching between the three waves, and on the
vicinity of cavity resonances for the signal and idler modes.
Theses frequencies are determined unambiguously when one knows the
values of two control parameters of the OPO, namely the crystal
temperature (which sets the value of the different indices) and
the cavity length (which determines the cavity resonance
conditions). Frequency degeneracy, \emph{i.e.}
$\omega_1=\omega_2=\omega_0/2$, occurs only accidentally since it
corresponds to a single point in the parameter space. It cannot be
achieved for a long time in real experimental conditions, as these
parameters drift in time. Furthermore, even when the device is
actively stabilized on the frequency degeneracy working point, the
signal and idler fields still undergo a phase diffusion
phenomenon, similar to the Schawlow-Townes effect in a laser
\cite{graham,courtois}, but acting on the difference between the
phases of the signal and idler modes in the case of the type II
OPO. As a result, the output field
polarization direction slowly drifts with time.\\

In the context of quantum information and generation of EPR correlated
bright beams, where both amplitude and phase correlations are involved,
phase locking is interesting since it allows a much simpler measurement of
amplitude \emph{and} phase quantum correlations between the signal and
idler beams \cite{part2} even above threshold : the measurement of
intensity quantum correlations between the signal and idler modes can
be done even with non-frequency degenerate beams \cite{mertz} but the
measurement of phase correlations makes it necessary to use a local
oscillator.  Thus a phase reference is defined and signal and idler
must be stable compared to this reference, which is not the case in a
regular, above threshold, type II OPO.

A few years ago, Wong \emph{et al.}  had the idea of achieving the
frequency degenerate operation at the output of a type-II OPO by
introducing a linear coupling between the signal and idler fields.
This coupling was made by way of a birefringent quarter-wave plate
placed inside the OPO linear cavity which couples the two
orthogonally polarized signal and idler waves. In this way, they
generated intense and stable frequency degenerate signal and idler
beams\cite{Wong98}. The theoretical model described in reference
\cite{fabre99} was able to account for the main features of this
phenomenon, but, for the sake of simplicity, it was made for a
ring cavity, for a small angle between the crystal neutral axes
and the birefringent plate neutral axes, and without any
phase-shifts introduced by the reflection on the cavity mirrors or
by non perfect phase matching.

Most experiments use linear cavities, whereas most theoretical
treatments assume ring cavities.  For scalar fields there is
almost no difference between the two configurations (the crystal
in the ring cavity being taken twice as long as the linear
cavity), if one neglects the cavity mirror differential
phase-shifts. This is no longer the case when polarization effects
are taken into account : in this case, a matrix formalism is
needed, and the exact succession of the different elements in the
cavity is now important, as they are described by non-commuting
matrices.  It is also interesting to examine the regime when the
birefringent plate angle is not limited to small values.  It seems
also important to take into account the mirror phase shifts, which
are known to induce a significant change in the phase matching
between the three waves and consequently in the oscillation
threshold of the linear cavity OPO \cite{debuisschert93}. The
purpose of the present paper is to introduce all these refinements
in the theoretical model introduced in \cite{fabre99}, and also to
discuss the properties of the phase-locked OPO in terms of the
actual control parameters of the device, which are the cavity
length and the crystal temperature.  This paper is followed by a
second one \cite{part2} in which the quantum fluctuations and
correlations between the signal and idler fields are determined
and studied in the same configuration.

In section (\ref{sec:systeme}), we introduce and describe the
behavior of the different elements placed inside the OPO cavity.
We then determine and discuss in section
(\ref{sec:etatstatanneau}) the steady-state regime in the ring
cavity case. Finally, in section (\ref{sec:etatstatlin}), we
examine and discuss the steady state regime in the linear cavity
case.

\section{Linear and nonlinear elements in the OPO cavity}
\label{sec:systeme}

We consider here a $\chi^{(2)}$ crystal with a type II phase matching,
 which means that the signal and idler fields have
orthogonal polarizations. The crystal length is $l$ and its
indices of refraction are $n_1$ and $n_2$ respectively for the
signal (ordinary) and idler (extraordinary) waves which are
supposed to be frequency degenerate. The non degenerate case will
be studied elsewhere \cite{nondeg}.

Assuming a small variation of the various field amplitudes inside the
nonlinear medium, which is quite reasonable in a c.w. OPO, one can
solve in an approximate way the propagation equations inside the
crystal, and obtain to the second order in the non-linearity, $g$~ :
\begin{equation}
\begin{array}{ccc}
A_0 (l) &=& A_0(0) - g \exp\left(-i \frac{\Delta k l}{2}\right)
\sinc \left(\frac{\Delta k l}{2}\right) A_1(0) A_2(0)  \\
&& \quad  - \frac{g^2}{2} f^\ast \left(\frac{\Delta k l}{2}
\right)\left( |A_1(0)|^2 + |A_2(0)|^2 \right) A_0(0) \\
A_1 (l) &=& A_1(0) + g \exp\left(i \frac{\Delta k l}{2}\right) \sinc
\left(\frac{\Delta k l}{2}\right)A_0(0) A_2^\ast(0)   \\
&&  \qquad + \frac{g^2}{2} f\left(\frac{\Delta k l}{2}\right)
\left(|A_0(0)|^2 -
  |A_2(0)|^2 \right)A_1(0)  \\
A_2 (l) &=& A_2(0) + g \exp\left(i \frac{\Delta k l}{2}\right) \sinc
\left(\frac{\Delta k l}{2}\right) A_0(0) A_1^\ast(0)  \\
&& \qquad + \frac{g^2}{2} f\left(\frac{\Delta k l}{2}\right)
\left(|A_0(0)|^2 -
  |A_1(0)|^2 \right)
A_2(0)
\label{eq:propagcristal}
\end{array}
\end{equation}
in which $A_0$ is the envelope amplitude and the $A_i, \, i=$ 1, 2 are
the envelope amplitudes of the interacting fields, assumed to be plane
waves, along the crystal axes ($C_1$ : ordinary wave, $C_2$ :
extraordinary wave); the envelopes are normalized in such a way that
$|A_i|^2$ gives the photon flow (photon.m$^{-2}$.s$^{-1}$); $g$ is the
nonlinear coupling coefficient given by
\begin{equation}
  \label{eq:gainparam}
  g = l \chi^{(2)} \sqrt{\frac{\hbar \omega_0 \omega_1 \omega_2}{2c^3
      \varepsilon_0 n_0 n_1 n_2}}
\end{equation}
and $f(x) = \frac{\exp(ix)}{ix} (\exp(ix)-\sinc (x))$. 
The crystal input-output equations are
then, when one expresses the pump field at the center of the
crystal~:

\begin{equation}
\begin{array}{ccc}
A_1 (l) &=& A_1(0) + g' A_0(\frac{l}{2}) A_2^\ast(0)  \\
A_2 (l) &=& A_2(0) + g' A_0(\frac{l}{2}) A_1^\ast(0)
\label{eq:simppropagcristal}
\end{array}
\end{equation}
where $g'=g \exp\left(i \frac{\Delta k l}{2}\right) \sinc
\left(\frac{\Delta k l}{2}\right)$. The advantage of this expression is
that it is valid to the second order in the non-linearity $g'$.

\begin{figure}
\centerline{\includegraphics[height=3cm]{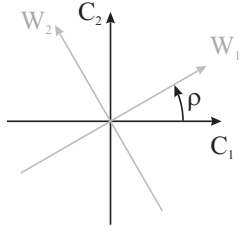}}
\caption{$\rho$ is the angle between the crystal's axes ($(C_1,\,
C_2)$, black lines) and the waveplate axes ($(W_1,\, W_2)$, grey
lines).\label{angles}}
\end{figure}

The second element in the cavity is the birefringent wave plate.
It has a thickness $e$ and its indices of refraction are $n_e$ and
$n_f$ at frequency $\omega_0/2$ respectively for the slow and fast
axes which make an angle $\rho$ with the crystal axes (see
fig.~\ref{angles}).

Its effect will be described in the Jones matrices formalism \cite{jones} in the
nonlinear crystal axes basis.  The transmission through the wave plate
can be written as the matrix~:
\begin{equation}
  \label{eq:propaglame}
  M = e^{ikne} \left( \begin{array}{cc} \alpha  &
      \epsilon \\ \epsilon & \alpha^\ast \end{array} \right)
\end{equation}
where
\begin{equation}
  n= \frac{n_s + n_f}{2}
\end{equation}
represents the mean index of refraction of the waveplate, and
\begin{eqnarray}
  \alpha &=& \cos \left(\frac{\Delta \phi}{2}\right) + i \cos (2\rho) \sin
  \left(\frac{\Delta \phi}{2}\right)\\
\epsilon &=& i \sin\left(\frac{\Delta \phi}{2}\right) \sin (2\rho)
\end{eqnarray}
where $\Delta \phi = k(n_s - n_f)e$ is the waveplate birefringent
phase-shift. Let us set $\alpha=\alpha_0 e^{i \psi}$ where
$(\alpha_0,\psi) \in \mathbb{R}^2$.  For the sake of simplicity, we
will assume that this plate has no effect on the pump field
polarization, i.e.  acts as a $\lambda$ waveplate at the pump
frequency.

\section{Ring cavity type II OPO}\label{sec:etatstatanneau}

\begin{figure}
\centerline{\includegraphics[width=.75\columnwidth]{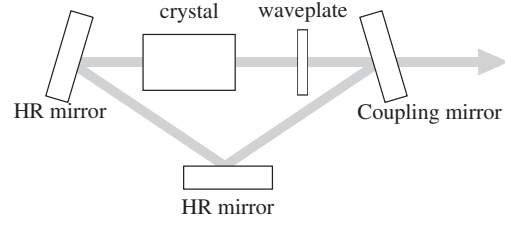}}
\caption{Set-up of the ring cavity; we consider only one direction
of propagation.\label{ringcav}}
\end{figure}

We assume in this section that the cavity has a ring shape
(fig.~\ref{ringcav}), and that the coupling mirror has large
reflection coefficients for signal and idler modes ($r_1$ and
$r_2$).  The moduli of the amplitude reflection coefficients of
the coupling mirror are taken equal for the signal and idler
modes~: $|r_1 |= |r_2|=r=1 - \kappa$, with $\kappa \ll 1$, so that
the transmission is $|t|\approx \sqrt{2\kappa}$. $\mu$ is the
round-trip loss coefficient for the signal and idler waves (due to
crystal absorption, surface scattering, other mirror finite
transmission ...), assumed to be small.  We define a generalized
reflection coefficient~: $r' = r (1- \mu) \approx 1 - ( \mu +
\kappa ) $. We will call $\zeta_1$ and $\zeta_2$ the phase-shifts
introduced by the reflection on the cavity mirrors for the signal
and idler waves so that $r_j = r \exp (i \zeta_j), \, j=1,2$. In
all the article, we do not take into account the resonance of the
pump mode~: all equations are given for the pump field inside the
crystal and we only calculate operating thresholds (not signal or
idler intensities) normalized to the intracavity pump threshold of
the OPO without the waveplate (standard OPO threshold),
$\sigma_0$. 
The free propagation length inside the cavity is
denoted $L$.

As the signal and idler fields are assumed to have the same
frequency, the birefringent plate and the non-linear crystal
couple the same fields, which are the ordinary and extraordinary
waves at frequency $\frac{\omega_0}{2}$, and only three complex
equations are needed to describe the system. From equations
(\ref{eq:simppropagcristal}) and (\ref{eq:propaglame}), one
readily obtains the following steady state equations for the field
amplitudes $A_1=A_1(0)$ and $A_2=A_2(0)$~:
\begin{equation}
\begin{array}{>{\displaystyle}c>{\displaystyle}c>{\displaystyle}c}
A_1 &=& r' \alpha_0 e^{i (\delta - \theta/2+ \psi)} \left( A_1 +
g' A_0 A_2^\ast \right) + \nonumber\\
&& \qquad r' \epsilon e^{i (\delta + \theta/2 )} \left( A_2 + g'
A_0
  A_1^\ast \right) \nonumber\\
A_2 &=& r' \alpha_0 e^{i (\delta + \theta/2 - \psi)} \left( A_2 +
g' A_0 A_1^\ast \right) +\label{eq:base} \\
&& \qquad r' \epsilon e^{i (\delta - \theta /2)} \left( A_1
  + g' A_0 A_2^\ast \right) \nonumber
\end{array}
\end{equation}
 where $\delta=\frac{\omega_0}{2c}\left(\frac{n}{2} e +
  \frac{n_1+n_2}{2} l +L\right) + \frac{\zeta_1 + \zeta_2}{2}$ is the
mean round-trip phase-shift, and $\theta
=\frac{\omega_0}{2c}(n_1-n_2) l + \zeta_1 - \zeta_2$ is the
birefringent phase-shift introduced by the non-linear crystal and
by the mirrors.

One immediately observes that these equations are not invariant
under the gauge transformation $A_1\longrightarrow A_1 e^{i
\varphi}$, $A_2\longrightarrow A_2 e^{-i \varphi}$, as is the case
for the usual equations of a non-degenerate OPO without
birefringent mixing. This implies that, unlike in the usual OPO,
the phases of the signal and idler amplitudes solutions of
equations (\ref{eq:base}), when they exist, are perfectly
determined : phase-locking has occurred between the two
oscillating modes, and there is no phase diffusion effect. This
phase-locking phenomenon is common to all linearly coupled
oscillators \cite{synchro}.

Since the effect of the different elements on the polarization is
described by matrices which do not commute, one expects that the
system depends on the plate position. However, it is
straightforward to show that exchanging the positions of the
waveplate and of the crystal amounts to a rotation of $\pi/2$ of
the crystal which is equivalent to exchanging indices 1 and 2.
This does not change the physics of the system so that we will
place ourselves in the case where the waveplate is located after
the crystal with respect to the input beam.

Equations (\ref{eq:base}) have been solved analytically in the small
angle regime $\rho \ll 1$ and for small cavity detunings and losses in
ref \cite{fabre99}.

We will present here the properties of the more complex analytical
solutions obtained without any approximations~: we will not give
the complicated expressions of the solutions, but instead give
plots of the most striking results. The exact expressions for the
different parameters in the case of a small angle are given in the
appendix.

\begin{figure}[h]
\begin{center}
\includegraphics[width=.75\columnwidth]{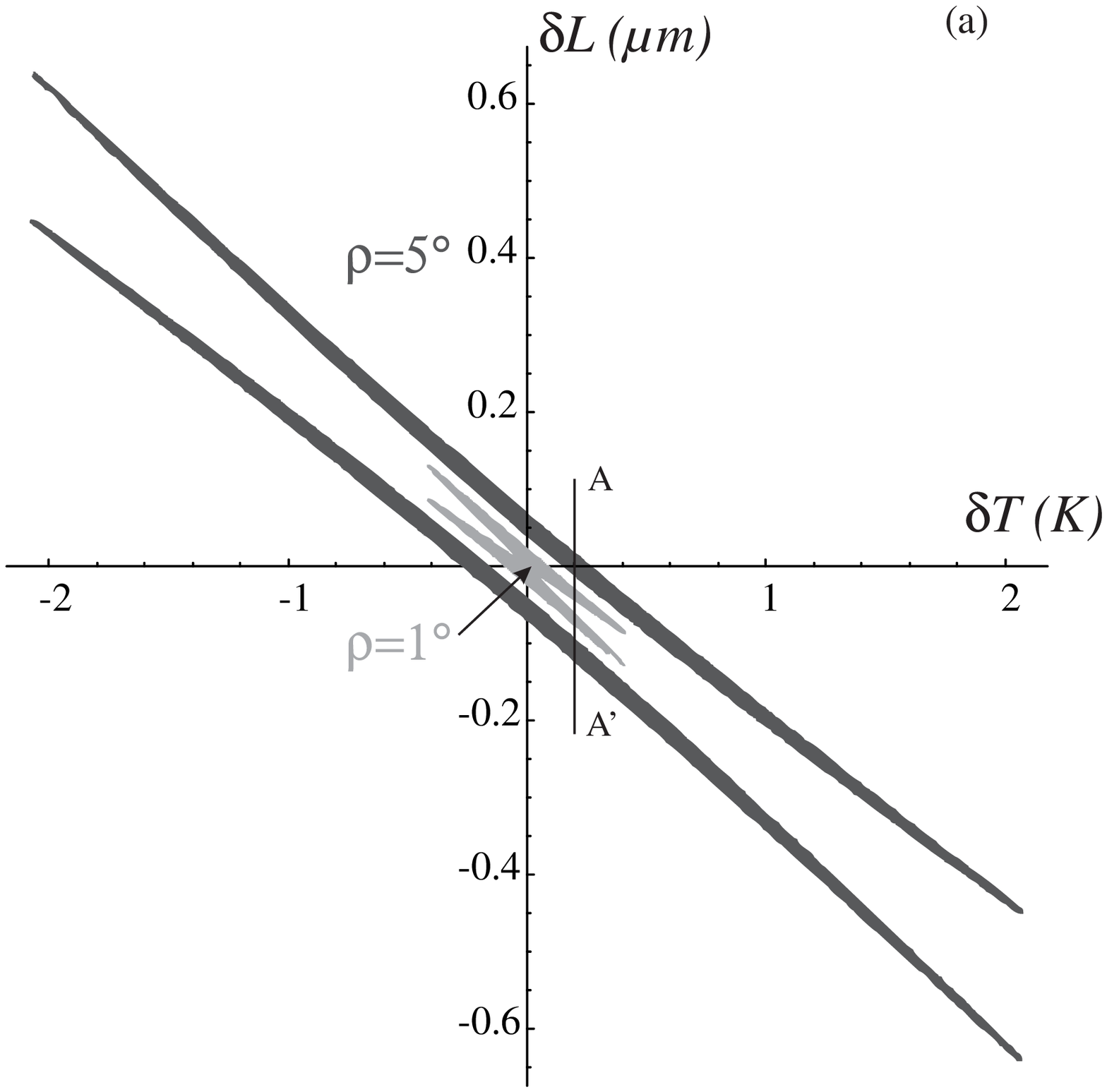}\\
\includegraphics[width=.75\columnwidth]{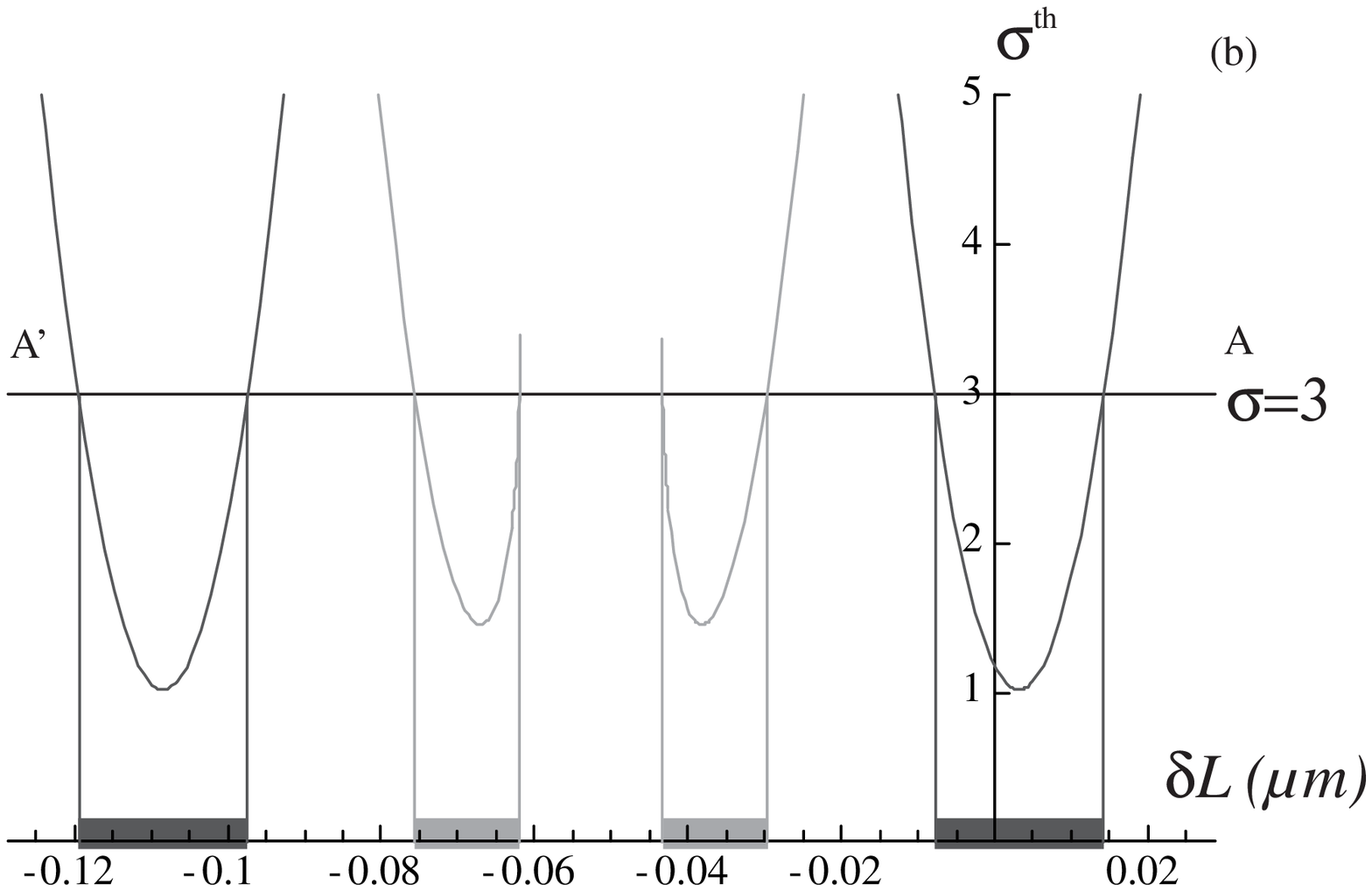}
\end{center}
 \caption{(a) : Locking zone as a function of the cavity length
 ($\delta L$) and of the crystal temperature ($\delta T$) for waveplate angle
 $\rho=1^\circ$ (light grey) and $\rho=5^\circ$ (dark grey).
$\sigma=3$. (b) : $\sigma^{th}$ as a function of the cavity length
($\delta L$) for $\delta T=0.2~K$. For $\sigma=3$, this
corresponds to the cross section AA' of the locking zone. $\Delta
\phi = \pi$. \label{lockzoneanneau}}
\end{figure}

The real and imaginary parts of the first two equations of
(\ref{eq:base}) form a set of two linear equations for the
amplitude and phase of the field enveloppes ${A_1,\, A_2}$. Thus,
one obtains a set of four linear equations with four variables. A
non-zero solution of this systems exists only when the
corresponding $4\times4$ determinant is zero.  This condition
gives a real equation for the system parameters, which is
fulfilled only in a specific operating range, or \textit{locking
zone}, for the self-phase-locked OPO.  In the locking zone, this
equation has two real solutions for the intracavity pump
intensity, corresponding to two possible regimes of the system
\cite{Wong98}. In this paper, we will focus our attention to the
regime of lower threshold. These solutions give the oscillation
threshold for the intracavity pump power as a function of the
crystal temperature, the cavity length and the waveplate angle. We
define $\sigma$ as the ratio of the intracavity pump power to
$\sigma_{0}$ and $\sigma^{th}$ as the ratio of the intracavity
pump power threshold to $\sigma_0$. If, for a given set of
parameters, $\sigma$ is larger than $\sigma^{th}$, one obtains
frequency degenerate oscillation. One can thus plot the values of
cavity length and crystal temperature for which $\sigma^{th}$ is
smaller than $\sigma$ so that there is degenerate oscillation.
Fig.(\ref{lockzoneanneau}-a) displays the locking zones for two
values of the wave plate angle $\rho$ as a function of $\delta
T=T-T_{deg}$ and $\delta L=L-L_{deg}$, where $T_{deg}$ is the
temperature for which the exact  frequency-degenerate operation
occurs without any birefringent coupling and and $L_{deg}$ the
corresponding cavity resonance length. The locking zone consists
of two surfaces which overlap for small values of $\rho$.
Fig.(\ref{lockzoneanneau}-b) shows the cross section AA' of the
locking zone for a given value of $\delta T$, that is
$\sigma^{th}$ as a function of $\delta L$. All curves in this
paper are plotted in the case of KTP for which the index of
refraction vary with the following dependance
\cite{cristallaser}~:
\begin{equation}
\frac{dn_1}{dT} = 1.3 \times 10^{-5}~K^{-1} \quad \mbox{and} \quad
\frac{dn_2}{dT} = 1.6 \times 10^{-5}~K^{-1}
\end{equation}

Fig.~(\ref{lockzoneanneau}-a) shows that the locking zone
extension increases as a function of $\rho$. However, the minimum
threshold does not increase with $\rho$ and a threshold equal to
the standard OPO threshold can always be found for $\delta T=0$.
For $\delta T\neq 0$, the minimum threshold as a function of
$\delta L$ is no longer equal to one (see
fig.~\ref{lockzoneanneau}, bottom).

\begin{figure}[h]
\centerline{\includegraphics[width=.75\columnwidth]{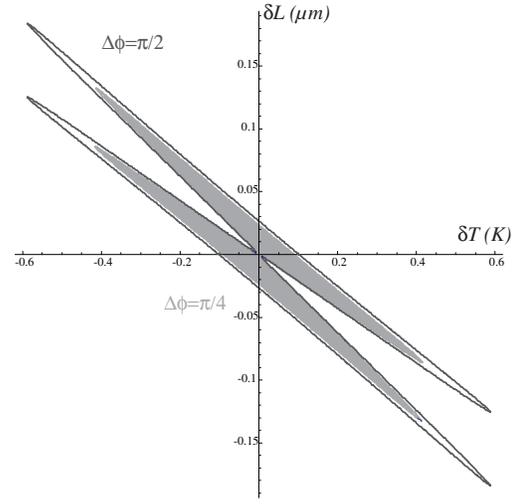}}
\caption{Locking zone as a function of cavity length and crystal
temperature. The thin dark grey line corresponds to a $\lambda/2$
waveplate while the light grey zone corresponds to a $\lambda/4$
waveplate. $\rho=5^\circ, \, \sigma=2$.\label{deltaphi}}
\end{figure}

For a given value of $\rho$, the coupling parameter $\epsilon =i
\sin(\frac{\Delta \phi}{2}) \sin (2\rho)$ is maximized for $\Delta
\phi=\pi$ that is for a $\lambda/2$ waveplate. As shown on fig.
(\ref{deltaphi}), a different value for $\Delta \phi$ will reduce
the locking zone extension but does not change the general shape.
In order to maximize $\epsilon$ and thus the locking zone
extension, one can set $\Delta \phi=\pi$ and $\rho=45^\circ$. In
this case, the locking zone is infinite, in practice only limited
by the phase matching.

\begin{figure}[h]
\centerline{\includegraphics[width=0.75\columnwidth]{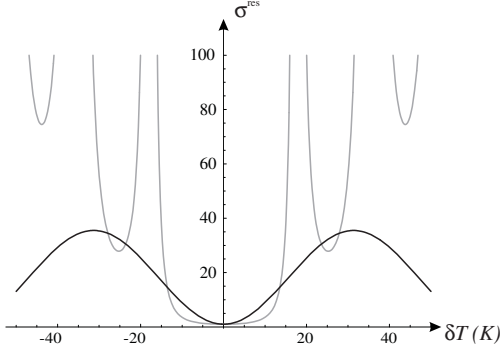}}
\caption{Normalized threshold on resonance $\sigma^{res}$ as a
function of the temperature for $\rho=30^\circ$. The black curve
corresponds to the result obtained without taking into account the
phase matching multiplied by 50 for readability of the figure. The
grey curve is plotted taking into account the phase
matching.\label{phmismatch}}
\end{figure}

As the locking zone depends on the temperature, it may be
important to take into account the phase matching. However for
small values of the waveplate angle, the locking zone extension in
$\delta T$ is small so that the effect of $delta k \neq 0$ remains
small. As $\rho$ is increased, this effect becomes noticeable and
limits effectively the extension of the locking zone to a zone
$\delta T \approx 10~K$. We have plotted on
fig.~(\ref{phmismatch}) $\sigma^{res}$, the threshold on resonance
: it corresponds to the minimum value of $\sigma$ as a function of
$\delta L$ for a fixed value of $\delta T$. One notices on this
figure that $\sigma^{res}$ is periodic in $\delta T$ if one does
not take into account the phase matching : this is due to the
periodicity in temperature of the crystal birefringence. When one
takes into account the phase matching, this periodicity disappears
(grey curve).

For realistic parameters, such as $R=90 \%$ and $\sigma=2$, the
transverse width of the locking zone is $\Delta L \approx \lambda
/\mathcal F \approx 10~nm$ where $\mathcal F$ is the cavity
finesse and $\Delta T \approx 50~mK$. These values give the
conditions on the length and temperature control loops to remain
within the locking zone. These constraints are compatible with the
current performances of length and temperature controls.

\section{Linear cavity type II OPO}
\label{sec:etatstatlin}

In this section, we study the linear cavity case which is actually
used in most experiments. We show here that the linear and ring
cavity OPO have different behaviors when one takes into account
the reflection phaseshifts on the cavity mirrors for the different
interacting waves.

\begin{figure}
\centerline{\includegraphics[width=0.75\columnwidth]{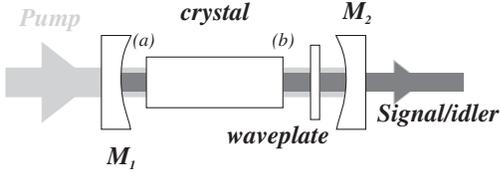}}
\caption{Set-up of the linear cavity type II OPO \label{lincav}}
\end{figure}

 One mirror, M$_1$ is highly reflective for
signal and idler and serves as a coupling mirror for the pump
while the other mirror, M$_2$ is highly reflective for the pump
and serves as a coupling mirror for signal and idler
(fig~\ref{lincav}).
The phase of the reflection coefficient for signal and idler are
taken equal.

We redefine the waveplate coupling constants since the signal and idler beams pass
two times in the waveplate\footnote{the free propagation and reflection on the
coupling mirror simply shift the two waves by the same phase which does not change
the effect of the waveplate}~:
\begin{eqnarray}
\alpha &=&\cos ( \Delta \phi )+i\sin (\Delta \phi )\cos (2\rho ) = \alpha
_{0}e^{i\psi }\\
\epsilon &=&i\sin (\Delta \phi )\sin (2\rho )
\end{eqnarray}
$\Delta \phi/2$ being replaced by $\Delta \phi$.

As mentioned in the introduction and in the previous section, the
linear cavity OPO has distinct features when compared to the ring
cavity while the triple resonance does not change the behavior of
the system. In the linear cavity the beams undergo two
interactions per round-trip. As the phase is important in a
parametric interaction the phase-shift between signal and idler
and the pump beam between the two nonlinear interactions in the
crystal must be taken into account. The equations for the field
enveloppes at face (a) of the crystal can be written to the first
order in $g'$~:
\begin{equation}
\begin{array}{rcl}
 A_{1}&=& \alpha _{0}r^{\prime } e^{i(\delta  - \delta'
)}[ A_1 +(1+e^{i\xi })g' A_{0}A_{2}^{\ast }]\\
&&  \qquad+\epsilon r^{\prime } e^{i \delta} [A_{2}+(1-e^{i\xi }) g' A_{0}A_{1}^{\ast } ]\\
A_{2} &=&  \alpha _{0}r^{\prime } e^{ i(\delta + \delta')}[A_2 +
(1+e^{i\xi }) g' A_{0} A_{1}^{\ast } ] \\ &&  \qquad + \epsilon
r^{\prime }e^{i \delta } [A_{1}+(1-e^{i\xi })gA_{0}A_{2}^{\ast }]
\end{array}\label{eq:linearcav}
\end{equation}

where
\begin{eqnarray}
\delta &=&\frac{\omega _{0}}{2c}(2ne+2\bar{n}l+2L)+\zeta_1+\zeta_2\\
\delta'&=&\theta -\psi\\
\delta _{0}&=&\frac{\omega _{0}}{c}(2n_{0}l+2L) + \zeta_{0}\\
\xi &=&\frac{\omega _{0}}{2c}(2n_{0}-2 \bar n)l- \frac{\omega _{0}}{c}n (2e) +
\zeta_0 - 2 \zeta_2
\end{eqnarray}
$2L$ is the total round-trip free propagation length. $\bar n$ and
$\theta$ have been defined in section \ref{sec:etatstatanneau}.

\begin{figure}
\centerline{\includegraphics[width=.75\columnwidth]{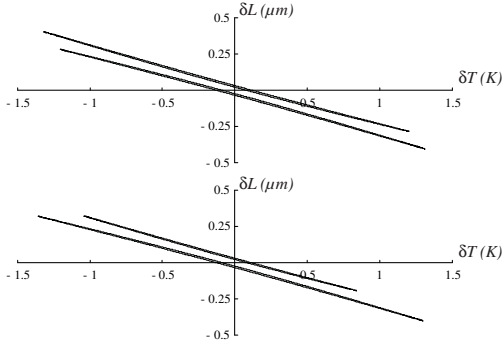}}
\caption{Locking zone as a function of cavity length and crystal
  temperature for two values of $\xi$~: top $\xi=0$, bottom
  $\xi=\pi/4$. The other values are the same : $\rho=5^\circ$,
  $\sigma=3$.\label{dissym}}
\end{figure}

One sees on the first two equations of expression
\ref{eq:linearcav} that when the phase-shift $\xi$ is taken equal
to 0, the equations are similar to the ring cavity case
\footnote{when one neglects the
  second order term in $\epsilon g$ if $\rho$ is taken to be small},
but with a crystal of double length (factor $2g'$).  This is no longer
the case when this parameter is changed. A non-zero value of $\xi$ has
been shown to increase the threshold of a standard OPO by a
significant amount \cite{debuisschert93}. In the case of a linear
cavity with a birefringent element, a dissymmetry appears in the
equations due to the terms $1 \pm e^{i\xi}$. Fig.\ref{dissym} shows an
example of the results obtained : one notices the dissymmetry between
the two locking zones.

\begin{figure}
\centerline{\includegraphics[width=0.75\columnwidth]{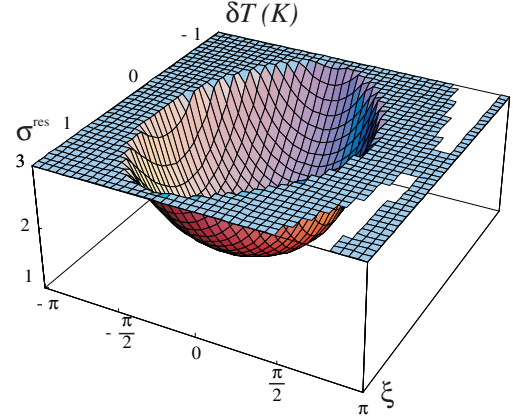}}
\caption{\label{sigma_xi_T_5}Normalized threshold on resonance
$\sigma_{res}$ as a function of the crystal temperature $\delta T$
and of the phase-shift $\xi$ for $\rho=5^\circ$. Unshaded surfaces
correspond to values of $(\delta T,\delta L)$ where frequency
degenerate operation is not possible.}
\end{figure}

Fig. \ref{sigma_xi_T_5} presents the value of the threshold on
resonance, $\sigma^{res}$ as a function of the temperature $\delta
T$ and the phase-shift $\xi$ for $\rho=5^\circ$. One observes that
this threshold is no longer obtained for $\delta T=0$ as is the
case for a ring cavity. For small values of $\rho$ and $\xi$,
$\sigma^{res}$ remains reasonable ($\sigma^{res} < 3$) inside a
temperature range of approximately $1~K$. This value is small
compared to the pure phase matching temperature range of $15~K$.
However, as $\xi$ increases and goes to $\pi$, $\sigma^{res}$
diverges~: as $\xi$ is fixed by the exact mirror dielectric
structure, it is not adjustable experimentally (except by changing
the mirrors) : this can be a severe limitation to operation of the
phase-locked OPO for small values of the waveplate angle $\rho$.

\begin{figure}
\centerline{\includegraphics[width=0.75\columnwidth]{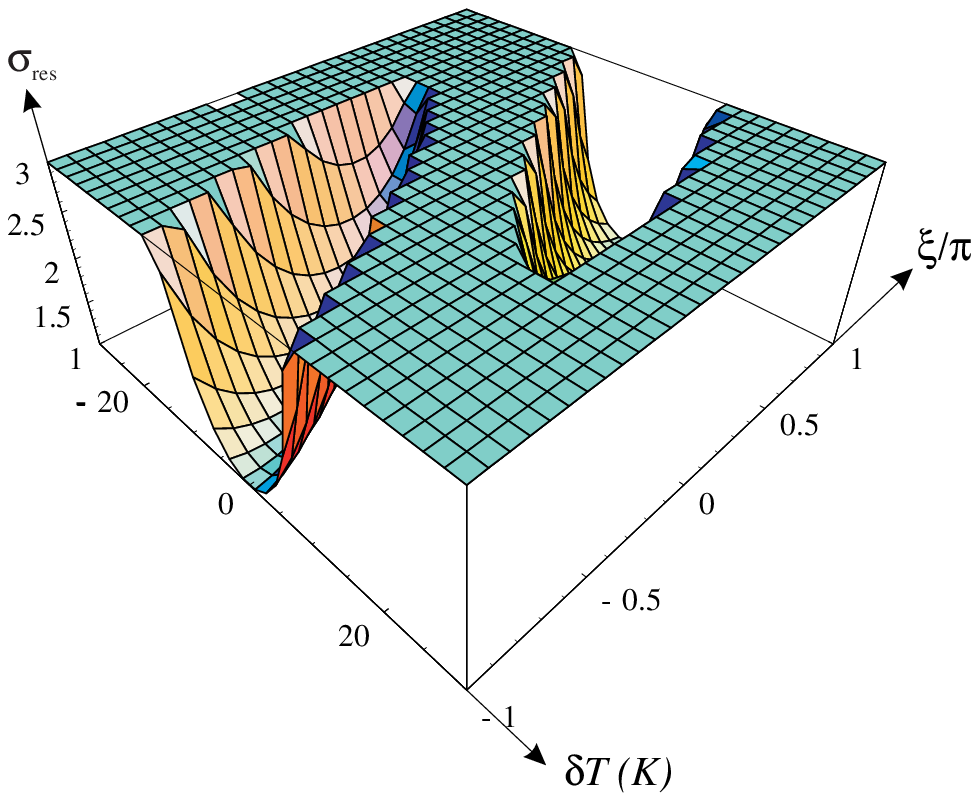}}
\centerline{\includegraphics[width=0.75\columnwidth]{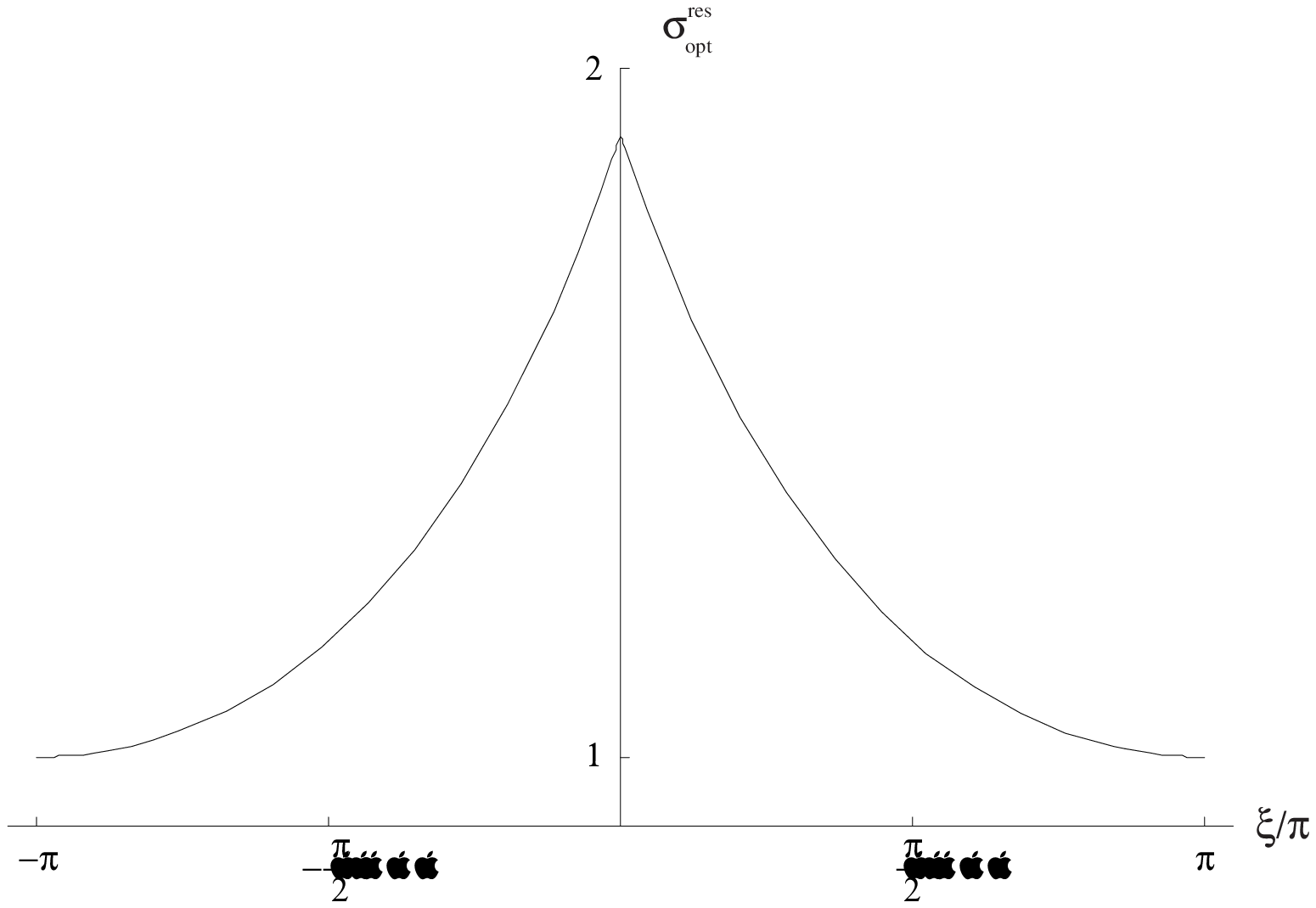}}
\caption{\label{sigma_xi_T_45}Normalized threshold on resonance
$\sigma_{res}$ as a function of the crystal temperature $\delta T$
and of the phase-shift $\xi$ for $\rho=45^\circ$ (top). Same curve
optimized for $\delta T$ (bottom)}
\end{figure}

When $\rho$ is increased, the locking zone size increases and a
larger range of temperature can be used with a low threshold.
Figure \ref{sigma_xi_T_45} shows the behavior of the normalized
threshold on resonance $\sigma_{res}$ as a function of $\xi$ and
$\delta T$ for $\rho=45^\circ$. In this case, the minimum value of
$\sigma_{res}$ is obtained for $\xi=\pi$ and $\delta T=0$. When
$\xi$ is lowered to $0$, $\sigma_{res}$  increases. The maximum
value of $\sigma_{res}$ is obtained for $\xi=0$ and two values of
the temperature : it is equal to 1.92 times the standard OPO
threshold. This increase by a factor 1.92 is also found for
$\xi=\pi$ in the case of the standard OPO \cite{debuisschert93}.

\section{Conclusion}

We have studied a system composed of an Optical Parametric
Oscillator containing a birefringent waveplate inside the optical
cavity. As shown previously \cite{Wong98,fabre99}, this system
allows phase locking of the signal and idler fields. We have
obtained equations that are valid for all wave plate angles as
well as in different cavity configurations, namely ring or linear
cavities. We have shown that the zone where phase locking occurs
can be described by the cavity length and the crystal temperature
and consists of two zones. As the waveplate angle is increased,
the size of the locking zone increases.  The optimal configuration
is obtained by inserting a $\lambda/2$ waveplate in a ring cavity
or a $\lambda/4$ waveplate in a linear cavity with a $45^\circ$
angle with respect to the crystal's axis. In the case of a ring
cavity, the minimum threshold is obtained for a temperature such
that the crystal birefringence compensates all the other
birefringence in the cavity (waveplate and mirrors) and is equal
to the standard OPO threshold. The effect of phase mismatch
between the three waves is small for small values of the waveplate
angle since the locking zone extension in temperature is small. As
$\rho$ is increased, the effect of phase mismatch becomes
noticeable and limits in practice the extension of the locking
zone. In a linear cavity, the mirrors phase-shift modifies the
minimum threshold which becomes dependent on the waveplate angle
and can become twice as large as the standard OPO threshold. This
increase is known even in standard OPOs but a linear cavity is
usually chosen for experimental reasons (losses, mechanical
stability\dots). In both cases (standard and self-phase-locked
OPO), this increase is accompanied of a shift in the optimal
crystal temperature which may be large and must be taken into
account to operate the OPO at low threshold.

\begin{acknowledgement}
Laboratoire Kastler-Brossel, of the Ecole Normale Sup\'{e}rieure
and the Universit\'{e} Pierre et Marie Curie, is associated with
the Centre National de la Recherche Scientifique.\\
This work was supported by European Community Project QUICOV IST-1999-13071\\
T. Coudreau is also at the P\^ole Mat{\'e}riaux et Ph{\'e}nom{\`e}nes
Quantiques FR CNRS 2437, Universit{\'e} Denis Diderot, 2, Place
Jussieu, 75251 Paris cedex 05, France
\end{acknowledgement} %

\section*{Appendix}

We give here the exact expression for the lower oscillation
threshold in the case of a ring cavity~:
\begin{equation}
I^{th} = \frac{u - \sqrt v}{g^{\prime 2} r^{\prime 2}}
\end{equation}
with
\begin{eqnarray}
u &=& \epsilon ^2 + r'^2 -
    2r'\alpha_0
\cos (\delta )
     \cos \left(\frac{\theta}{2}  - 2\psi \right) \nonumber\\
     &&\qquad  +
    {\alpha_0}^2\cos \left(\theta  - 2\psi \right) \\
v &=&
  \left[ r^{\prime2} + \epsilon_0^2 - 2 r^\prime \alpha_0 \cos (\delta )
\cos \left(\frac{\theta}{2}  -\psi \right)  + \alpha_0^2
\cos^2 (\theta  - 2\psi ) \right]^2 \nonumber\\
&& -1 - r^{\prime4} - 2 r^{\prime2}\alpha_0^2  - 2 r^\prime
\Bigg\{ r^\prime\cos (2\delta ) +\alpha_0  \times  \nonumber\\ &&
\left[ r^\prime \alpha_0 \cos (\theta - 2\psi ) -2 \left( 1 +
r^{\prime2} \right) \cos (\delta ) \cos \left(\frac{\theta}{2} -
\psi\right)\right] \Bigg\}
\end{eqnarray}
with the parameters defined in the text.

\end{document}